\documentclass[aps,prl,twocolumn,superscriptaddress,floatfix,longbibliography]{revtex4-1}
\usepackage[dvipsnames]{xcolor}
\usepackage{amssymb,amsmath,amsfonts,bm,slashed,soul,ragged2e,graphicx,epstopdf,hyperref,array}
\usepackage[utf8]{inputenc}

\usepackage{colortbl,color,epsfig,subfigure,caption}
\enlargethispage{\baselineskip}
\definecolor{steelblue}{RGB}{25,25,112}
\definecolor{dullblue}{rgb}{0,0.298,0.49}
\definecolor{darkred}{rgb}{0.545,0,0}
\definecolor{blue2}{cmyk}{1, 0.1, 0.1, 0}
\hypersetup{colorlinks,linkcolor={darkred},citecolor={blue},urlcolor={dullblue}}
\DeclareCaptionJustification{justified}{\justifying}
\everymath{\displaystyle} 

\usepackage{comment}
\usepackage{braket}
\usepackage{amsmath}
\usepackage{soul}
\usepackage{makecell}
\usepackage{verbatim}

\usepackage{orcidlink}

\newcommand{\beq}{\begin{equation}}
\newcommand{\eeq}{\end{equation}}
\newcommand{\bea}{\begin{eqnarray}}
\newcommand{\eea}{\end{eqnarray}}

\newcommand{\gsim}{\lower.7ex\hbox{$\;\stackrel{\textstyle>}{\sim}\;$}}
\newcommand{\lsim}{\lower.7ex\hbox{$\;\stackrel{\textstyle<}{\sim}\;$}}

\newcommand{\be}{\begin{equation}}
\newcommand{\ee}{\end{equation}}
\newcommand{\ba}{\begin{eqnarray}}
\newcommand{\ea}{\end{eqnarray}}

\RequirePackage[normalem]{ulem}

\begin{document}

\title{Search for Ultralight  Axion Dark Matter with a Levitated Ferromagnetic Torsional Oscillator}

\author{Chunlong Li}
\affiliation{School of Physics, Hefei University of Technology, Hefei 230601, Anhui, China}
\author{Yiwei Huang}
\affiliation{School of Physics, Hefei University of Technology, Hefei 230601, Anhui, China}
\author{Shien Yang}
\affiliation{School of Physics, Hefei University of Technology, Hefei 230601, Anhui, China}
\author{Yichong Ren}
\email{renyichong@outlook.com \,(Corresponding author)}
\affiliation{School of Physics, Hefei University of Technology, Hefei 230601, Anhui, China}
\author{Yu Zhang\,\orcidlink{0000-0001-9415-8252}}
\email{dayu@hfut.edu.cn \,(Corresponding author)}
\affiliation{School of Physics, Hefei University of Technology, Hefei 230601, Anhui, China}
\author{Peiran Yin}
\affiliation{National Laboratory of Solid State Microstructures and Department of Physics, Nanjing University, Nanjing 210093, Jiangsu, China}
\author{Pu Huang}
\affiliation{National Laboratory of Solid State Microstructures and Department of Physics, Nanjing University, Nanjing 210093, Jiangsu, China}
\author{Fei Xue\,\orcidlink{0000-0002-5681-229X}}
\email{xfei.xue@hfut.edu.cn \,(Corresponding author)}
\affiliation{School of Physics, Hefei University of Technology, Hefei 230601, Anhui, China}

\begin{abstract}
We present a search for ultralight axion dark matter coupled to electron spins using a levitated ferromagnetic torsional oscillator (FMTO). This platform directly measures axion-induced torques on a macroscopic spin-polarized body, combining large spin density with strong mechanical isolation to probe magnetic fluctuations below 10\,Hz while suppressing gradient-field noise. In a first implementation, the experiment yielded $18\,000~\mathrm{s}$ of analyzable data at room temperature under high vacuum with optical readout and triple-layer magnetic shielding. A likelihood-based statistical framework, incorporating stochastic fluctuations in the axion-field amplitude, was used to evaluate the data. No excess consistent with an axion-induced pseudo-magnetic field was observed near $2\times10^{-14}\,\mathrm{eV}$. For possible shielding-induced signal attenuation, we quantify its effect and report both the uncorrected ($g_{aee}\lesssim10^{-7}$) and attenuation-corrected ($g_{aee}\lesssim6\times10^{-5}$) 90\%-C.L. limits on axion-electron coupling. Looking ahead, improvements guided by both noise-budget analysis and shielding-attenuation considerations—including optimized levitation geometry, cryogenic operation, and superconducting shielding—are expected to boost sensitivity by multiple orders of magnitude.
\end{abstract}

\date{\today}

\maketitle

\noindent{\it Introduction.}
---Astrophysical and cosmological observations have provided overwhelming evidence for the existence of dark matter (DM) \cite{Planck:2018vyg}. However, the origin of DM particles remains one of the most puzzling mysteries in physics. Except for the existing gravitational interaction of DM, other important fundamental particle properties such as its mass, spin and  potential interactions with other particles remain elusive \cite{Bertone:2016nfn, Freese:2017idy}. Among the most competitive DM candidates, the ultralight axions or axion-like particles (ALPs) have attracted particular attention \cite{Duffy:2009ig, Chadha-Day:2021szb}. These particles emerge as compelling candidates due to their potential to address multiple theoretical challenges simultaneously. Originally, the axion was a scheme proposed by Peccei, Quinn and others \cite{Peccei:1977hh, Peccei:1977ur, Weinberg:1977ma, Wilczek:1977pj} to resolve the strong charge-parity problem in quantum chromodynamics (QCD), but many theories in particle physics predict the existence of ALPs \cite{Graham:2015ouw, Irastorza:2018dyq}. Unlike traditional dark matter candidates, like Weakly Interacting Massive Particles (WIMPs), ultralight axion or axion-like dark matter (hereafter referred to as axion DM) exhibits wave-like behavior on astronomical scales \cite{Hui:2016ltb,Hui:2021tkt}, which could influence structure formation and alleviate small-scale problems in cosmological simulations \cite{Bullock:2017xww, Tulin:2017ara, Marsh:2015xka}. Their non-gravitational interactions with Standard Model particles, though weak, offer potential pathways for experimental detection, making them a vibrant focus of theoretical and experimental research \cite{Graham:2015ouw}.

Axions can interact with Standard Model fields through several distinct couplings, which give rise to different experimental search strategies \cite{Irastorza:2018dyq}. The coupling to photons allows for interconversion between axions and electromagnetic waves in strong magnetic fields, forming the basis of haloscope and helioscope experiments \cite{ADMX:2021nhd,Adair:2022rtw,Brubaker:2016ktl,Lazarus:1992ry,CAST:2011rjr,TASTE:2017pdv,IAXO:2019mpb}, as well as light-shining-through-a-wall searches \cite{Cameron:1993mr,Robilliard:2007bq,GammeVT-969:2007pci,Afanasev:2008jt,Ehret:2010mh,OSQAR:2015qdv,Betz:2013dza}. The coupling to gluons induces an oscillating electric dipole moment (EDM) in nucleons, which can be probed by precision measurements using storage rings or atomic and molecular EDM experiments \cite{Chang:2017ruk,JEDI:2022hxa,Flambaum:2019emh}. Meanwhile, the coupling to fermion spins, often referred to as the ``axion wind" interaction, leads to spin-precession effects in polarized samples, providing the foundation for a series of nuclear magnetic resonance (NMR)–based and comagnetometer experiments \cite{JacksonKimball:2017elr,Wei:2023rzs,Bloch:2021vnn,Jiang:2021dby,Zhang:2025zcq,Wei:2023rzs}. Together, these approaches form complementary avenues toward probing the wide mass range of axion dark matter.

Motivated by the pursuit of higher-precision spin-based measurements, macroscopic platforms with large spin polarization and strong mechanical isolation—such as levitated ferromagnetic systems—have emerged as promising tools for exploring new physics \cite{Ahrens:2024yzo,Vinante:2021fpj,Kimball:2016zza}. In conventional comagnetometers, coherence times are limited by atomic diffusion and collisional broadening \cite{PhysRevLett.95.230801}. By contrast, levitated ferromagnetic oscillators behave as macroscopic rigid bodies with minimal spin-relaxation pathways, allowing pioneering theoretical and experimental studies to achieve energy resolutions well below $\hbar$ \cite{Ahrens:2024yzo,Vinante:2021fpj} and to surpass the standard quantum limit for magnetometry with independent spins—arising from spin-projection noise—through collective spin precession in macroscopic samples \cite{Kimball:2016zza}. The macroscopic torsional motion of the ferromagnet converts weak magnetic-field–induced torques directly into measurable angular displacements, scaling linearly with total spin polarization rather than its gradient, and thereby bypassing magnetic-field gradient noise that typically constrains precession-based systems \cite{pang2023comprehensive}. Furthermore, diamagnetic levitation of the ferromagnet provides strong mechanical isolation, suppressing vibrational coupling and Johnson noise from supports \cite{chen2022diamagnetic}. These combined advantages—including high spin density, long coherence times, mechanical isolation, and linear torque-to-displacement transduction—make levitated ferromagnetic systems promising platforms for searches of dark matter and fifth forces \cite{Fadeev:2020xzw,Kalia:2024eml,Gao:2025eye}.

We report on a millimeter-scale ferromagnetic torsional oscillator (FMTO) operating at room temperature, aimed at probing ultralight axion dark matter via axion–electron interactions. We first characterize the FMTO system, including its construction and calibration procedures. We then develop a statistical framework to extract potential axion-induced torque signals and derive constraints on the axion–electron coupling coefficient $g_{aee}$. Finally, we propose strategies for future upgrades aimed at enhancing sensitivity.

\noindent{\it FMTO magnetometer for axion wind.}
---The axion is a well-motivated candidate for ultralight DM, with mass $m_a$ in the range $10^{-19}\,\mathrm{eV} \lesssim m_a c^2 \lesssim 1\,\mathrm{eV}$ \cite{Ferreira:2020fam}. To account for the entire local dark matter density, $\rho_a \approx 0.4\,\mathrm{GeV}c^{-2}\mathrm{cm}^{-3}$ \cite{Catena:2009mf,Read:2014qva}, such small masses imply a macroscopic occupation number within a single de Broglie volume $\lambda_{\mathrm{dB}}^3$. Consequently, DM behaves as a classical wave rather than as individual particles on macroscopic scales. These axion waves oscillate coherently at an angular frequency $\omega_a \approx m_a c^2/\hbar$, with a small fractional frequency spread $\Delta\omega \approx (v_0/c)^2 \omega_a$ set by the virial velocity $v_0 \approx 10^{-3} c$ \cite{Evans:2018bqy}. This frequency spread defines coherent domains traveling at an average velocity $v_0$, with coherence maintained over a timescale $T_{\mathrm{coh}} \equiv \lambda_{\mathrm{dB}}/v_0$. Within this regime, the axion field at spacetime point $(t, \vec{x})$ can be expressed as
\begin{align}
    a(\vec{x},t)=\frac{\hbar\sqrt{\rho_a}}{m_a c^2} \alpha \cos\left(\omega(t)t+\vec{k}(t)\cdot\vec{x}+\phi \right).
    \label{unaf}
\end{align}
Here $\vec{k} \approx m_a \vec{v}/\hbar$ is the momentum, and $\alpha$ and $\phi$ are random variables arising from sampling the axion field over time intervals shorter than the coherence time $T_{\mathrm{coh}}$. In the nonrelativistic regime, the axion–electron coupling manifests as an effective magnetic field $\vec{B}_a$ acting on electron magnetic moment $\mu_e$, giving an interaction Hamiltonian $\mathcal{H} = -\vec{\mu}_e\cdot\vec{B}_a$. The component of $\vec{B}_a$ along the detector’s sensitive-direction unit vector $\vec{\xi}$ is
\begin{align}
    B_a = \frac{\hbar}{c^3} \frac{g_{aee}}{2m_e \gamma_e} \sqrt{\rho_a} \alpha v_{\text{obs}}\cos\psi(t) \cos\left(\omega_a t+\phi \right),
    \label{effBa}
\end{align}
where $m_e$ and $\gamma_e$ denote the electron mass and gyromagnetic ratio, respectively. $v_{\mathrm{obs}} = 220\ \mathrm{km/s}$ is the average axion velocity, set by the motion of the Sun through the Galaxy. $\psi(t)$ denotes the time-dependent angle between $\vec{B}_a$ and $\hat{\xi}$. Its time dependence arises from the Earth’s motion in the Galactic frame, with the dominant modulation set by the Earth’s rotation \cite{Wu:2019exd}. To enhance the signal of $B_a$, one may increase the number of sensing electrons $N_e$, thereby boosting the total magnetic moment $\mu = N_e \mu_e$. Ferromagnets, which provide a high density of electron spins with macroscopic magnetization, thus serve as an effective probe \cite{Kimball:2016zza}.

\begin{figure}[htbp]
    \centering
    \includegraphics[width=0.47\textwidth]{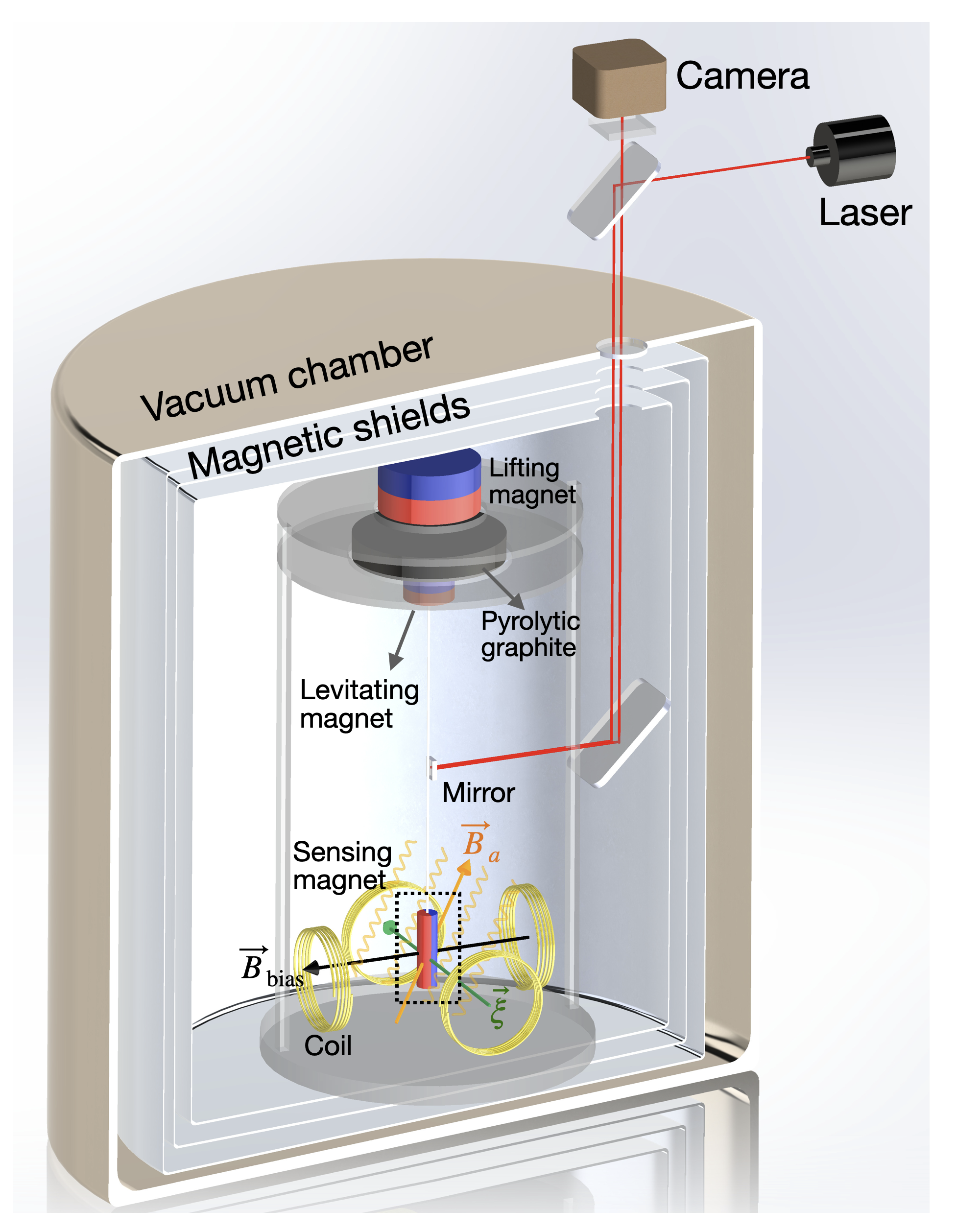}
    \caption{Schematic of the FMTO system. The system consists of a levitated magnet, a sensing magnet, and a connecting Teflon rod. A pyrolytic graphite disk provides stable levitation of the assembly. The motion of the FMTO is monitored using a camera combined with a centroid-tracking algorithm. A pair of coils generates the bias magnetic field $\vec{B}_{\mathrm{bias}}$, inducing the torsional eigenmode of the sensing magnet around its symmetry axis, while the axis perpendicular to this coil pair defines the detector’s sensitive axis $\vec{\xi}$ for the axion-wind effective magnetic field $\vec{B}_a$. Another pair of coils, aligned along $\vec{\xi}$, is used to apply calibration signals to characterize the system response.}
    \label{fig:setup}
\end{figure}

In the previous work, we designed a ferromagnetic torsional oscillator (FMTO) for detecting weak magnetic fields \cite{Ren}. Its detection sensitivity and frequency coverage also make it suitable as an ultralight axion dark matter detector. The experimental setup is depicted in Fig.\,\ref{fig:setup}. The weak magnetic field probe is a cylindrical ferromagnet made of NdFeB, with a diameter of 1 mm and a height of 20 mm, whose magnetization direction $\vec{\mu}$ is along the diameter. The supporting structure of the ferromagnetic probe consists of a Teflon rod attached to a levitating magnet, which is separated from the upper lifting magnet by a pyrolytic graphite disk. The lifting magnet holds the entire oscillator, providing a levitated state. The readout apparatus is implemented using the optical lever method, where the mirror on the Teflon rod deflects the laser, thereby mapping the torsional motion of the FMTO to the translation of the laser spot on the image plane, which is captured by the camera and resolved by a centroid algorithm. The FMTO is placed inside a three-layer magnetic shield and a vacuum chamber at room temperature. Windows in the shields and vacuum chamber allow the laser beam to enter and exit. Thermal and vibrational isolations is also implemented throughout the system, including for the FMTO, magnetic shields, and measurement optics.

The sensing ferromagnet is polarized by a DC-driven coil pair, enabling torsional motion about its symmetry axis. In the frequency domain, the angular displacement is $\theta(\omega)=\chi(\omega)\mu B_s(\omega)/I$ with mechanical susceptibility $\chi(\omega) = 1 / \left(\omega_0^2-\omega^2+i \omega \omega_0 / Q\right)$, the quality factor $Q=\omega_0/\gamma$, and the resonant angular frequency $\omega_0=\sqrt{k/I}$. The measurable magnetic-field component $B_s$ is the projection of the field along the sensitive direction $\vec{\xi}$, and can also be generated by a second coil pair to provide a calibrated test signal for determining $\omega_0$ and $Q$.

The raw data are the displacement of the optical spot on the imaging plane, recorded in pixels as Y(t), which traces the motion of the sensing magnet. We work in Fourier space by analyzing the magnetic PSD $S_B(f)$, related to the displacement PSD via $S_B(f)=\mathcal{T}(f)S_Y(f)$. The transfer function $\mathcal{T}(f)$ is determined by applying an off-resonant calibration field of known amplitude $B_{\mathrm{cal}}$ at frequency $f_{\mathrm{cal}}$. Using the resulting displacement amplitude $Y_{\mathrm{cal}}$ together with the independently calibrated $\chi(f)$, the transfer function $\mathcal{T}(f)$ is then obtained. In Table\,\ref{sigma}, we list the calibrated parameters and their associated uncertainties.
\begin{table}[h]
\centering
\setlength{\tabcolsep}{16pt} 
\begin{tabular}{cc}
\hline
\hline
Parameters & Values \\
\hline
$f_0$ & $5.00\pm 0.01\,\mathrm{Hz}$ \\
$Q$ & $39.2\pm 0.4$ \\
$f_{\mathrm{cal}}$ & $0.1\pm 10^{-6}\,\mathrm{Hz}$ \\
$B_{\mathrm{cal}}$ & $114\pm 17\,\mathrm{pT}$ \\
$Y_{\mathrm{cal}}$ & $(8.5\pm 0.5) \times 10^{-2}\,\mathrm{pixels}$ \\ [0.5em]

$S_0$ & $0.477\,\mathrm{pT^2/Hz}$ \\
$S_1$ & $0.14\,\mathrm{pT^2/Hz^5}$ \\ [0.5em]

$I$ & $2.6\times 10^{-3}\,\mathrm{g\cdot cm^2}$ \\
$N_e$ & $9.4\times 10^{20}$ \\
\hline
\hline
\end{tabular}
\caption{Uncertainties of the experimentally calibrated parameters and fitted values of $S_0$ and $S_1$ extracted from the magnetic PSD. Also included for reference are the oscillator’s moment of inertia $I$ about the sensing-magnet symmetry axis and the number of electrons $N_e$ in the sensing magnet.}
\label{sigma}
\end{table}

\begin{figure}[htbp]
    \centering
    \includegraphics[width=0.47\textwidth]{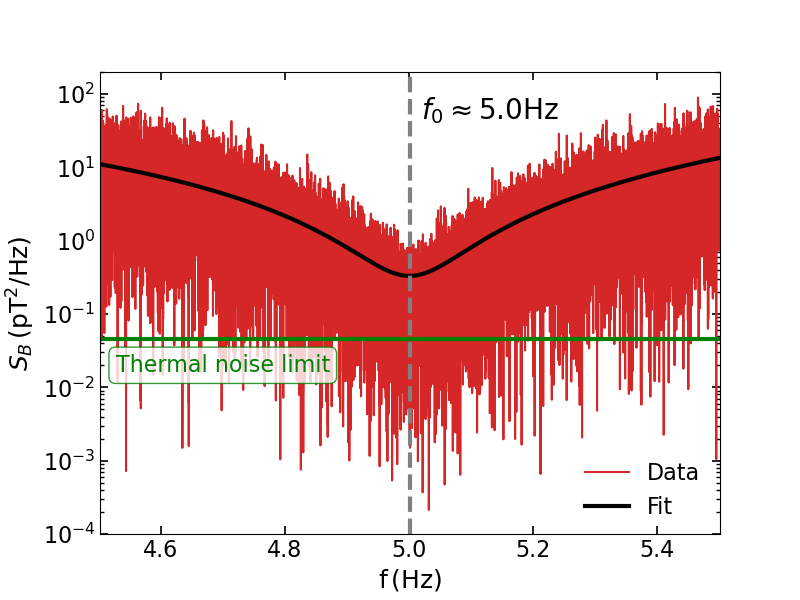}
    \caption{The magnetic PSD $S_B(f)$ measured around the resonance point and the fitted magnetic noise baseline. The vertical line highlights the resonance frequency $f_0\approx 5\text{Hz}$ of the torsional mode.}
    \label{fig:psd}
\end{figure}

\noindent{\it Data analysis and results.}
---The ultralight axion dark matter signal appears as a single-frequency peak in the magnetic PSD, with this frequency bin containing the Compton frequency $f_c \approx m_a c^2 / (2\pi\hbar)$. To search for this signal, data acquisition proceeded continuously for approximately 6 hours. A continuous segment of duration $T_{\text{mea}} = 18\,000~\mathrm{s}$ was selected for data analysis, after removing an initial unstable interval during which the laser spot drifted outside the imaging plane. Fourier transformation yields the magnetic PSD $S_B(f)$ near the resonance frequency $f_0$, as shown in Fig.\,\ref{fig:psd}. The data were fit to the total expected magnetic noise via $\bar{S}_{B}(f)=S_0 + \left\vert \chi(f) \right\vert^{-2}S_1$ (black curve), with the theoretical thermal noise at room temperature indicated for comparison (green line).

To quantify the significance of the signal relative to the expected magnetic noise, the key quantity is the power excess in each frequency bin, defined as the PSD value in that bin normalized by the expected noise baseline, $p(f)\equiv S_B(f)/\bar{S}_B(f)$. Accordingly, the theoretical signal PSD $S_A$ needs to be normalized to $p_A$, i.e.,
\begin{align}
    p_A(f, \alpha) &\equiv \frac{S_A(\alpha)}{\bar{S}_B(f)}=A(f) \alpha^2, \nonumber \\
    A(f) &\equiv \frac{\hbar^2}{c^6} \frac{g_{aee}^2}{8 m_e^2\gamma_e^2}\rho_a v_{obs}^2 \cos^2\psi T_{\mathrm{mea}} / \bar{S}_B(f).
    \label{ps}
\end{align}
Since the Fourier-transform sampling time is shorter than a sidereal day, the time dependence of $\psi(t)$ can be neglected, and we use the average value $\cos^2\psi=0.061$ over the sampling time (see Supplemental Material). To perform our inference on $p_A$, we follow a frequentist, likelihood-based approach similar to Refs.\,\cite{Amaral:2024tjg,Amaral:2024rbj}. The likelihood of a measured $p$ in any frequency bin is a non-central $\chi^2$ with two degrees of freedom and a non-centrality parameter $p_A$ \cite{1975ApJS...29..285G}. To account for the stochastic amplitude $\alpha$ of the axion dark matter field, we marginalize this likelihood by integrating over its probability distribution. The result is an exponential likelihood with inverse scale $1+2A$.

To test whether a nonzero $p_{A}$ is present in a given frequency bin, we construct a two–sided test statistic (TS) using the log–likelihood ratio formalism \cite{Cowan:2010js}. We first determine the TS distribution under the null hypothesis (no axion signal) via Monte Carlo simulations, from which we obtain the local p-value $p_{\rm val}^0$ for each frequency bin. The smallest observed local p-value is $p_{\rm val}^0 = 3.3\times10^{-5}$. However, because the analysis involves $N_p=18{,}000$ statistically independent frequency bins, this excess must be corrected for the look-elsewhere effect. The corresponding global p-value is estimated as $N_p p_{\rm val}^0 = 0.59$, which corresponds to a global significance of only $0.53\sigma$. Thus, the excess is fully consistent with a statistical fluctuation, and no axion signal can be claimed.

\begin{figure}[htbp]
    \centering
    \includegraphics[width=0.47\textwidth]{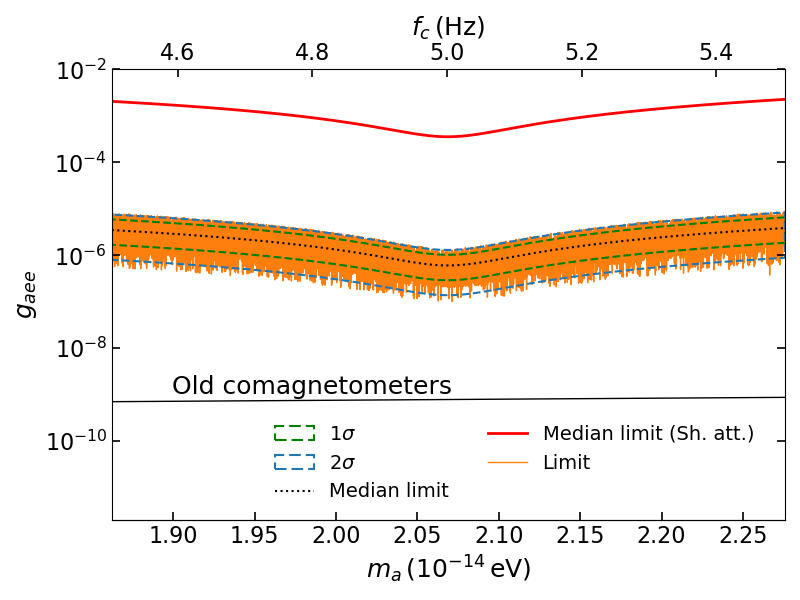}
    \caption{The 90\% confidence level limits on the axion-electron coupling strength $g_{aee}$ with axion mass $m_a$ (bottom axis) and Compton frequency $f_c$ (top axis). Shown are the data-driven limits derived from the measurements presented in Fig.\,\ref{fig:psd}, as well as the median limit and $1\sigma/2\sigma$ bands derived from our simulations. Also shown are the existing limits from the old comagnetometers \cite{Bloch:2019lcy}. ``Shielding attenuated (Sh. att.)'' denotes the inclusion of a potential attenuation factor on the axion-induced field $B_a$ due to the innermost magnetic shielding layer.
    }
    \label{fig:constraints}
\end{figure}

We place 90\% confidence level limits by solving for the axion–electron coupling constant $g_{aee}$ from the p-value $p_{\rm val}^{A} = 0.1$, with the results shown in Fig.\,\ref{fig:constraints}. To assess whether our limit is consistent with that from background-only data, we also derive the expected $1\sigma$, $2\sigma$ and median limits by simulating background-only pseudo-data and repeating the above procedure for each dataset, which are shown as dashed and dotted curve in Fig.\,\ref{fig:constraints}. We find excellent agreement between the results from our simulation and our derived data-driven limit, which closely follows the median limit and lies well within the $2\sigma$ band. Our best constraint is $g_{aee}\lesssim 10^{-7}$, occurring at the DM mass $m_a c^2\approx 2.07\times 10^{-14}$\ eV. This limit is not as stringent as those set by the noble-alkali comagnetometer \cite{Bloch:2019lcy}, shown as the black line in Fig.\,\ref{fig:constraints}. For other ultralight mass ranges of axion DM, a torsion pendulum experiment sets a limit of $3\times10^{-10}$ below $2\ \mathrm{mHz}$ \cite{Terrano:2019clh}. In the $10\text{–}100\ \mathrm{kHz}$ band, the fermionic-axion interferometer provides constraints at the level of $10^{-4}$–$10^{-7}$ \cite{Crescini:2023zyl}. At much higher frequencies near $10\ \mathrm{GHz}$, the ferromagnetic haloscope achieves a limit of about $2\times10^{-11}$ \cite{QUAX:2020adt}.

Ref.\,\cite{JacksonKimball:2016wzv} suggested that the innermost layer of magnetic shielding used in detection may substantially suppress the axion-induced signal, particularly for searches targeting the axion–electron coupling. We note that the existing limits on ultralight axion–electron interactions obtained using magnetic-shielded setups—specifically the old comagnetometers \cite{Bloch:2019lcy} and torsion pendulum \cite{Terrano:2019clh}—did not account for this potential attenuation of the axion-induced signal. Following the arguments of Ref.\,\cite{JacksonKimball:2016wzv}, we evaluate the maximal attenuation factor $G$ arising from the innermost permalloy shield in our setup and obtain $G \approx 1.7\times10^{-3}$ (see Supplemental Material). In Fig.\,\ref{fig:constraints}, we indicate the resulting impact on the limits with the red curve.

\noindent{\it Future Prospects.}
---Our result represents the first data-driven constraint on the coupling between ultralight axion dark matter and electrons obtained using a levitated ferromagnetic sensor. To advance further, an experimentally guided strategy for system upgrades based on the potential shielding-induced attenuation and the noise-budget analysis is required.

At the current stage, the experimental sensitivity has not yet reached the intrinsic thermal noise (Fig.\,\ref{fig:psd}). The dominant limitation arises from mechanical vibrations, which excite the pendulum mode at $1.5$ Hz, slightly elevating the noise floor near the 5 Hz torsional resonance and restricting the usable bandwidth in the off-resonant region. The first key upgrade is to improve the vibration-isolation system, which is expected to suppress these vibration-induced excitations, enhance the sensitivity by an order of magnitude, and extend the measurement bandwidth to approximately 1–20 Hz. In addition, mechanical vibrations also induce a slow drift of the resonance frequency, which limits the maximum achievable integration time. To mitigate this, we combine the upgraded isolation system with drift correction via reference calibration signals, automated re-locking of the optical lever, and active vibration damping \cite{xie2025frequency}, which is expected to extend the feasible measurement duration from 6 hours to about one month.

\begin{figure}[htbp]
    \centering
    \includegraphics[width=0.47\textwidth]{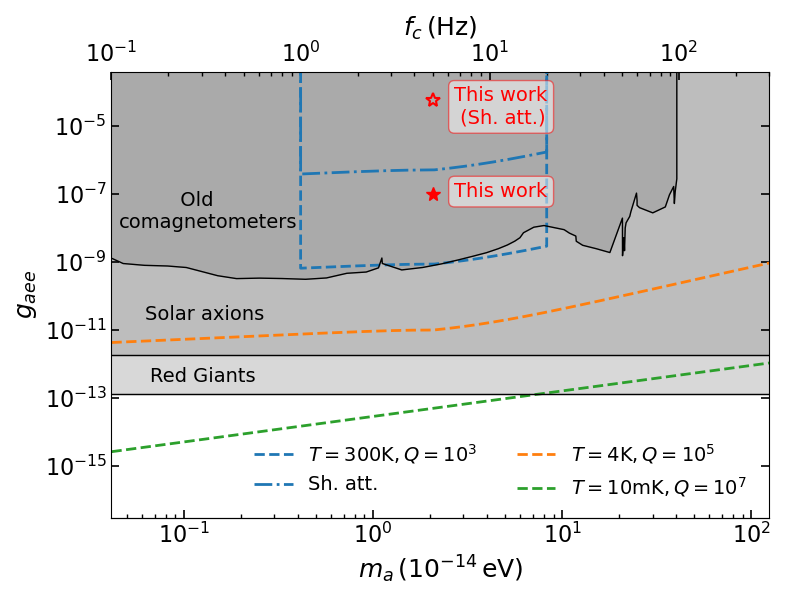}
    \caption{Projected sensitivity of the levitated FMTO to the axion–electron coupling $g_{aee}$ versus axion mass $m_a$, assuming a one-month integration time and the sidereal-day–averaged factor $\cos^2\psi = 0.22$. The star marks the most stringent limit from the current room-temperature, vibration-limited experiment ($\cos^2\psi = 0.061$). Curves show projections for optimized FMTO configurations at various temperatures and quality factors $Q$, including improved room-temperature designs with/without conservative shielding-attenuation corrections, and cryogenic setups featuring enhanced vibration isolation, higher $Q$, and superconducting shielding. Existing limits from comagnetometers \cite{Bloch:2019lcy}, XENONnT solar-axion searches \cite{XENON:2022ltv}, and red-giant cooling \cite{Capozzi:2020cbu} are also shown, highlighting that upgraded FMTO experiments can approach or compete with current laboratory and astrophysical constraints.}
    \label{fig:prospect}
\end{figure}

Once the thermal-noise limit is reached, further gains require suppressing the thermal noise itself. In parallel, the readout will be refined to ensure that its noise contribution remains subdominant relative to the thermal noise, both at resonance and throughout the sub-resonant band (see Supplemental Material). Since the thermal-noise PSD scales as $S_B^{\mathrm{th}} \propto I T/(Q \mu^2)$, raising the oscillator quality factor to $Q \sim 10^4$ yields a direct reduction in magnetic thermal noise. This can be achieved by optimizing the levitation geometry and replacing conventional diamagnetic components with low-dissipation composite materials \cite{Ji:2025yvn,Xie:2023pvn,Xiong:2025ppq}. In addition, improved diamagnetic levitation enables reducing the radius of the levitated magnet while enlarging and elongating the sensing magnet, thereby lowering $I/\mu^2$ by an order of magnitude. The projected sensitivity after these upgrades—shown as the blue dashed (dash-dotted) curve in Fig.\,\ref{fig:prospect} for the case without (with) shielding-induced attenuation—demonstrates that an optimized room-temperature experiment can reach the existing limits on ultralight axion–electron couplings set by comagnetometer measurements.

To further enhance the sensitivity and eliminate potential shielding-induced attenuation, the system can be cooled using a liquid-helium cryostat to reach 4 K. Such cryogenic operation substantially reduces the thermal-noise level and increases the mechanical quality factor of the levitated oscillator \cite{Vinante:2020zjt}. At these low temperatures, superconducting magnetic shielding becomes available, which—because it operates via electron orbital angular momentum rather than spin—does not attenuate axion-induced signals \cite{JacksonKimball:2016wzv}. Combined with a fully optimized vibration-isolation system, the cryogenic setup can also greatly expand the accessible measurement bandwidth. The resulting projected sensitivity is shown by the orange dashed curve in Fig.\,\ref{fig:prospect}. Such low-temperature, superconducting experiments can achieve sensitivities approaching astrophysical bounds. By further employing a dilution refrigerator to reach millikelvin temperatures, the most stringent limits in the sub-10 Hz low-frequency region can be realized, as shown by the green dashed curve in Fig.\,\ref{fig:prospect}. The most challenging parameters, $T = 10\ \mathrm{mK}$ and $Q = 10^7$, have been essentially realized or closely approached in separate experimental tests \cite{Vinante:2020zjt,Fuchs:2023ajk}.

\noindent{\it Conclusions.}
---We have demonstrated a levitated ferromagnetic torsional oscillator as a new platform for probing ultralight axion dark matter through its coupling to electron spins. This marks the first application of a macroscopically polarized spin system—levitated ferromagnets—to detect axion dark matter. Using a likelihood-based inference framework that incorporates stochastic fluctuations in the axion-field amplitude, we analyzed $18\,000~\mathrm{s}$ of room-temperature data and found no evidence of an axion-induced pseudo-magnetic field near $2\times10^{-14}\ \mathrm{eV}$. To assess the impact of possible signal attenuation by high-permeability shielding, we quantified its effect and reported both the uncorrected and attenuation-corrected 90\%-C.L. limits, yielding $g_{aee}\lesssim10^{-7}$ and $g_{aee}\lesssim6\times10^{-5}$, respectively. Building on these results, we outline realistic paths for significant sensitivity gains—including optimized levitation geometry, cryogenic operation, and superconducting shielding—establishing a clear route for future FMTO-based searches to access previously unexplored regions of axion parameter space.

\hspace{5mm}
\begin{acknowledgements}

\noindent{\it Acknowledgements.} 
We would like to thank Ran Ding for useful discussions.
This work was supported by the National Natural Science Foundation of China (Grants Nos. 12447128, 12475106 and T2388102), the China Postdoctoral Science Foundation (Grants No. 2024M760720), the National Key R\&D Program of China (Grant No. 2024YFF0727902), and the Fundamental Research Funds for the Central Universities (Grant No. JZ2025HGTB0176).
\end{acknowledgements}

\bibliographystyle{JHEP}
\bibliography{prl.bib}

\appendix
\pagebreak
\widetext
\begin{center}
\textbf{\large Supplemental Materials: Search for Ultralight  Axion Dark Matter with a Levitated Ferromagnetic Torsional Oscillator}
\end{center}
\setcounter{equation}{0}
\setcounter{figure}{0}
\setcounter{table}{0}
\makeatletter
\renewcommand{\theequation}{S\arabic{equation}}
\renewcommand{\thefigure}{S\arabic{figure}}
\renewcommand{\bibnumfmt}[1]{[#1]}
\renewcommand{\citenumfont}[1]{#1}

\section{Ultralight Axion Dark Matter}

Ultralight axion dark matter can be modeled as a classical field of mass $m_a$, and is treated as a superposition of a large number of states, forming a wave with a coherence time $T_{\text{coh}}\approx 2\pi\times 10^6 \hbar/m_a c^2$. If the detector’s single integration time is shorter than $T_{\mathrm{coh}}$, the frequency splittings among different states cannot be resolved. However, their relative oscillation phases must still be taken into account \cite{Nakatsuka:2022gaf,centers2021stochastic}. In this case, the superposition of axions with a given momentum $\vec{k}$ can be written as \cite{Foster:2017hbq,Lisanti:2021vij}
\begin{align}
    a(\vec{x},t)=\frac{\hbar}{c^2}\sum_{j=1}^{N_a} \frac{\sqrt{2\rho_a/N_a}}{m_a} \cos(\omega(t) t + \vec{k}(t)\cdot \vec{x}+\phi_j),
    \label{af}
\end{align}
where $j \in \{1,2,\ldots, N_a\}$ labels a specific axion state and $N_a\gg 1$. The angular frequency is given by $\omega(t) \approx m_a(1 + v(t)^2/2)$, where $\vec{v}(t)$ represents the axion velocity relative to the detector, and the corresponding momentum is $\vec{k}(t) \approx m_a \vec{v}(t)$. The time dependence of $\vec{v}(t)$ arises from the relative motion between the terrestrial laboratory and the Galactic, such that $\vec{v}(t) = \vec{u} + \vec{v}_{\mathrm{obs}}(t)$, where $\vec{u}$ and $\vec{v}_{\mathrm{obs}}(t)$ are the axion and the terrestrial laboratory velocities in the Galactic frame, respectively. $\vec{v}_{\mathrm{obs}}$ is dominated by the sun's motion in the Galactic frame, with a magnitude of $v_{\mathrm{obs}}\approx 220 \mathrm{km/s}$. The sum over $j$ reflects the fact that there are $N_a$ axions with effectively indistinguishable velocities $\vec{u}$ but independent random phases $\phi_j$. The oscillatory contribution can then be expressed as
\begin{align}
    \sum_{j=1}^{N_a} \cos(\omega(t) t + \vec{k}(t)\cdot \vec{x}+\phi_j) = \mathrm{Re} \left( e^{i (\omega t + \vec{k}\cdot\vec{x} )} \sum_{j=1}^{N_a} e^{i\phi_j} \right).
    \label{sumosc}
\end{align}
The sum over random phases can be treated as a two-dimensional random walk \cite{Foster:2017hbq}, yielding
\begin{align}
    \sum_{j=1}^{N_a} e^{i\phi_j} = \alpha e^{i\phi},
    \label{sumpha}
\end{align}
and $\alpha$ and $\phi$ follow the probability distributions
\begin{align}
    P_{\alpha}(\alpha)=\alpha e^{-\frac{\alpha^2}{2}}, \quad P_{\phi}(\phi)=\frac{1}{2\pi}.
    \label{random}
\end{align}
Here we have absorbed the factor of $N_a$ into $\alpha$ via the rescaling $\alpha \rightarrow \alpha \sqrt{N_a/2}$. Combining Eqs.\,(\ref{af}), (\ref{sumosc}) and (\ref{sumpha}), the resulting axion field can be written as
\begin{align}
    a(\vec{x},t)=\frac{\hbar\sqrt{\rho_a}}{m_a c^2} \alpha \cos\left(\omega(t)t+\vec{k}(t)\cdot\vec{x}+\phi \right).
    \label{unaf}
\end{align}
It is straightforward to verify that the average of Eq.\,(\ref{unaf}) satisfies $\langle a(t)^2\rangle=\hbar^2\rho_a/(m_a^2 c^4)$, using the relation $\int \alpha_\lambda^2 P_{\alpha}(\alpha_\lambda)=2$.

\section{Pseudo-magnetic Signal}

To detect an oscillating axion field with a terrestrial sensor, one can search for its possible non-gravitational interactions with Standard Model particles. In this work, we focus on the coupling between axions and electron spins, described by the Lagrangian \cite{graham2013new}
\begin{align}
     \mathcal{L}= \frac{g_{aee}}{2m_e c} \partial _{\mu}a\left( \vec{r},t \right) \overline{\varPsi }_e\gamma^{\mu}\gamma_5\varPsi_e,
\end{align}
where $g_{aee}$ parameterizes the axion–electron spin coupling strength and $m_e$ is the electron's mass. In the non-relativistic limit, the corresponding interaction Hamiltonian takes the form
\begin{align}
    \mathcal{H}=-\vec{B}_a\cdot\vec{\mu}, \quad \vec{B}_a\equiv \frac{g_{aee} \nabla a}{2\hbar c m_e \gamma_e}, \quad \mu\equiv N_e\mu_e,
    \label{Hami}
\end{align}
where $\gamma_e = 2\mu_e/\hbar$ is the electron gyromagnetic ratio and $\mu_e$ is the Bohr magneton. Here $N_e$ is the total number of electrons participating in the sensing process, related to the total electron spin $\vec{S}$ via $N_e = 2S/\hbar$. Eq.\,(\ref{Hami}) indicates that the axion wind generates a torque on the spin ensemble, $\vec{\tau}_a = \vec{\mu} \times \vec{B}_a$, which provides the experimental handle for detection.

Assuming that the probe experiences an angular acceleration along $\hat{\epsilon}$ under the torque $\vec{\tau}_a$, the signal response can be expressed as $\vec{\tau}_a\cdot \hat{\epsilon} \propto (\hat{\mu}\times \vec{B}_a)\cdot \hat{\epsilon}$, where $\hat{\mu}$ denotes the direction of the total magnetic moment of the spin ensemble. Defining the effective sensitive direction as $\hat{\xi}\equiv \hat{\epsilon}\times \hat{\mu}$, the detector response becomes $\vec{\tau}_a\cdot \hat{\epsilon} \propto \vec{B}_a \cdot \hat{\xi}$. Since a portion of the axion velocity $\vec{v}$ arises from the random velocity $\vec{u}$, the total detector response must be averaged over all possible directions \cite{Garcon:2019inh}. Assuming an isotropic distribution of $\vec{u}$ in the Galactic frame, this angular average yields $\int (d\Omega/4\pi) \nabla a \cdot \hat{\xi} \propto v_{\text{obs}} \cos\psi(t)$, where $\psi(t)$ is the angle between the laboratory velocity $\vec{v}_{\text{obs}}$ and the detector’s sensitive axis $\hat{\xi}$. The axion-induced effective magnetic field projected along $\hat{\xi}$ simplifies to
\begin{align}
    B_a = \frac{\hbar}{c^3} \frac{g_{aee}}{2m_e \gamma_e} \sqrt{\rho_a} \alpha v_{\text{obs}}\cos\psi(t) \cos\left(m_a t+\phi \right),
    \label{effBa}
\end{align}
where the usual approximation $\omega\approx m_a$ is used since the kinetic term is negligible compared to the rest mass of the particle. For a series of time-domain signal $B_a(t_n)$ with $n = 0, 1, \ldots, N-1$, we can derive the one-sided power spectral density (PSD) using the discrete Fourier transform,
\begin{align}
    S_A(f)=2\frac{(\Delta t)^2}{T_{\text{mea}}} \left\vert \sum_{n=0}^{N-1} B_a(t_n) e^{i 2\pi f n \Delta t} \right\vert^2,
    \label{fourier}
\end{align}
where $N$ is the number of points sampled in the time domain, and $\Delta t \equiv T_{\mathrm{mea}}/N$ is the sampling interval. The factor of two accounts for the folding of the result from negative frequencies to positive frequencies to produce the one-sided periodogram. The resulting signal PSD is given by
\begin{align}
    S_A(\alpha)=\frac{\hbar^2}{c^6} \frac{g_{aee}^2}{8 m_e^2 \gamma_e^2}\rho_a \alpha^2 v_{obs}^2 \cos^2\psi \, T_{\mathrm{mea}}.
    \label{signalPSD}
\end{align}
Note that the signal PSD for $T_{\mathrm{mea}} < T_{\mathrm{coh}}$ has no dependence on the axion mass $m_a$.

\section{Time Modulation}

\begin{figure}[htbp]
    \centering
    \includegraphics[width=0.45\textwidth]{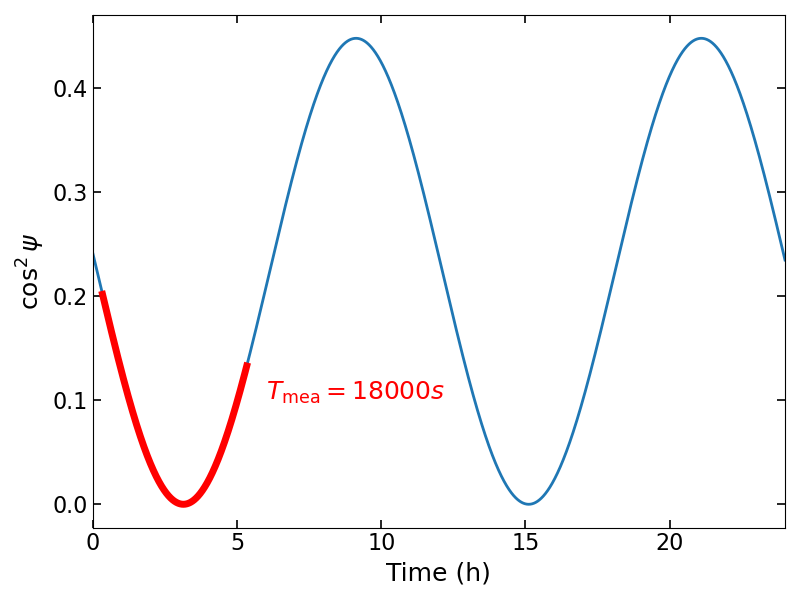}
    \caption{Variation of $\cos^2\psi$ over a sidereal day, with the red segment indicating the time window corresponding to the analyzed data.}
    \label{fig:cospsi2}
\end{figure}

Due to the Earth’s motion in the Galactic frame, the angle $\psi(t)$ between the axion wind and the detector’s sensitive axis varies in time, inducing a characteristic modulation of the axion-induced signal. This modulation is primarily generated by the Earth’s rotation and appears at the frequency of sidereal day $\Omega_d = 1/T_d$ \cite{Bandyopadhyay:2010zj}. In celestial coordinates, the time dependence of $\psi(t)$ is given by \cite{Wu:2019exd}
\begin{align}
    \cos\psi(t) = \cos \chi \sin \delta+\sin \chi \cos \delta \cos (\Omega_d t-\zeta),
    \label{psi}
\end{align}
where $\delta = -48^{\circ}$ and $\zeta = 138^{\circ}$ are the declination and right ascension of the axion velocity direction in celestial coordinates, respectively. The angle $\chi$ denotes the separation between $\hat{\xi}$ and the celestial north direction, given by $\cos\chi = \cos\gamma \cos\beta$, where $\gamma$ is the geographic latitude of the detector and $\beta$ is the angle between $\hat{\xi}$ and the geographic north. For our experiment, $\gamma=32^\circ$ and $\beta=90^\circ$. In Fig.\,\ref{fig:cospsi2}, we show the variation of $\cos\psi^2$ over one sidereal day. For experiments with a single integration time longer than $T_d$, this daily modulation leads to a spectral splitting of the dark-matter signal at frequencies $m_a c^2/(2\pi \hbar) \pm 1/T_d$. In the present experiment, however, the measurement time satisfies $T_{\mathrm{mea}}<T_d$, so the Fourier resolution is insufficient to resolve this splitting. We therefore use the time-averaged value over the measurement interval $T_{\mathrm{mea}}=18{,}000\,\mathrm{s}$, yielding $\cos^2\psi = 0.061$.

\section{Experimental Calibration}

\begin{figure}[htbp]
    \centering
    \includegraphics[width=0.45\textwidth]{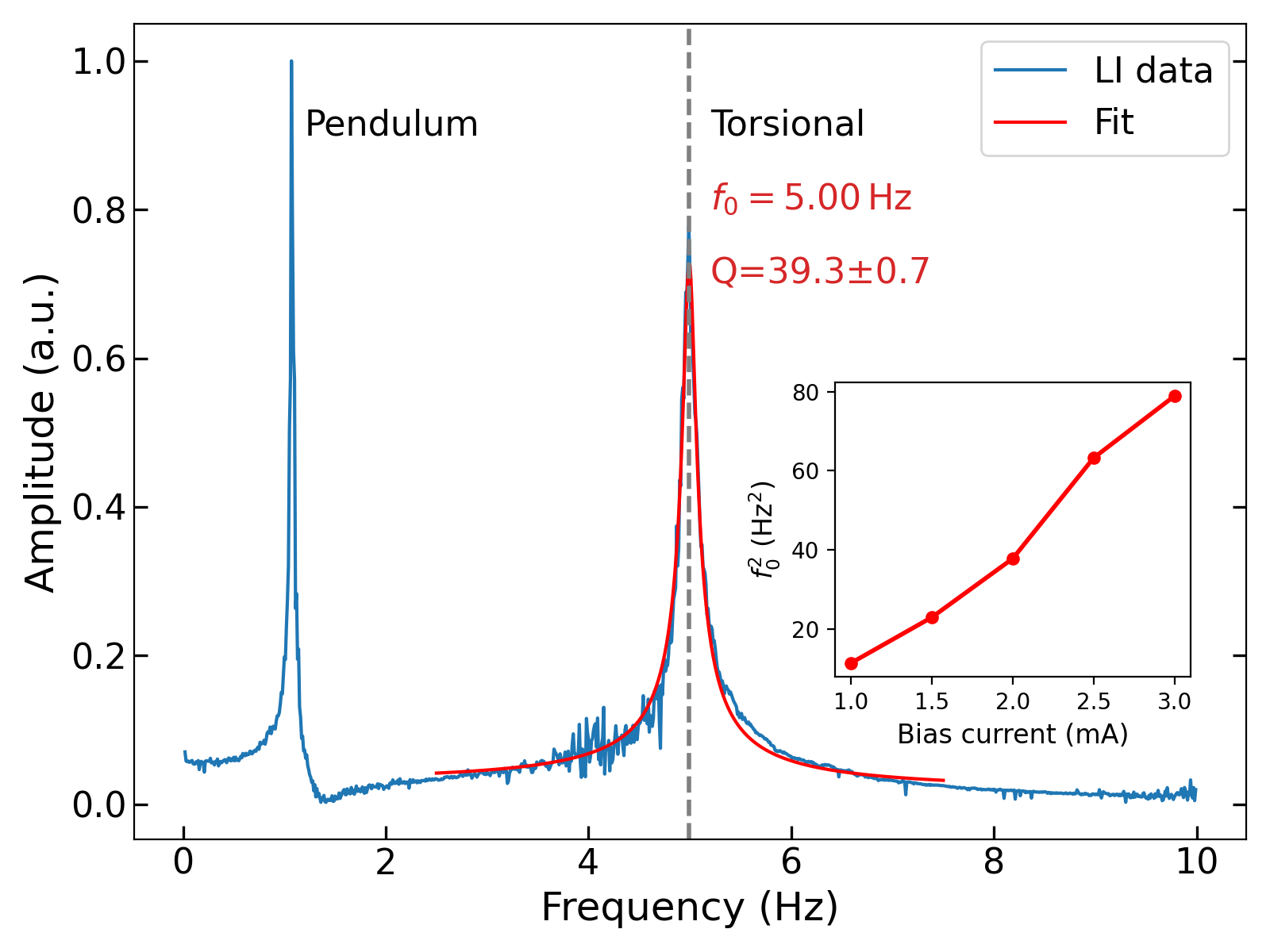}
    \includegraphics[width=0.45\textwidth]{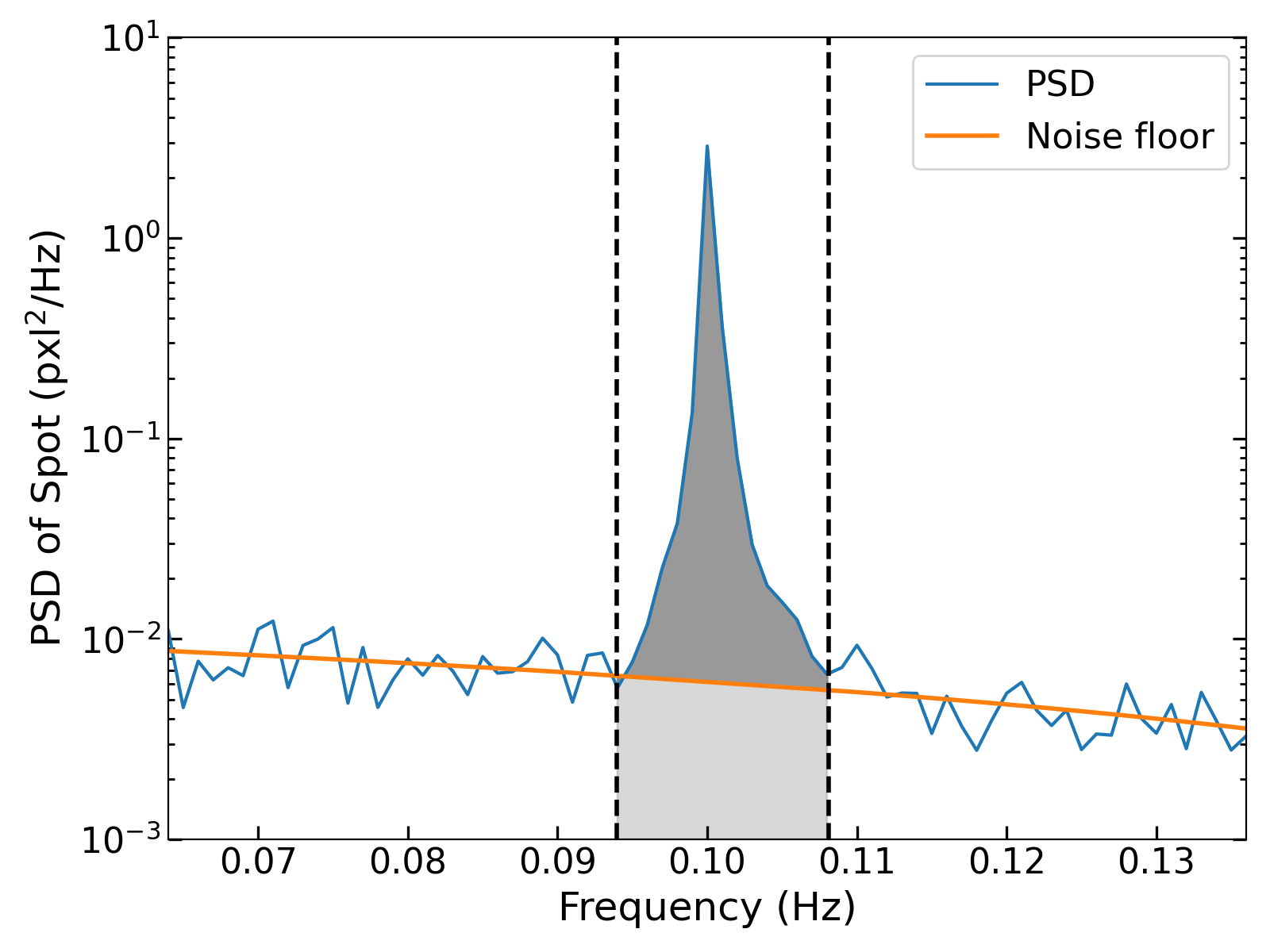}
    \caption{Left: Resonance peak and quality factor obtained via post-processing digital lock-in scanning. Two modes appear below 10 Hz, with the higher-frequency mode identified as the mode of torsional oscillation about the symmetry axis. The resonance frequency and $Q$ are extracted by Lorentzian fits. The inset shows the dependence of the squared torsional-mode frequency on the current used to generate the bias magnetic field. Right: Calibration of the transfer function from the magnetic field amplitude to the optical spot displacement. A reference field $B_{\mathrm{cal}}$ is applied at frequency $f_{\mathrm{cal}}$, and the corresponding displacement amplitude $Y_{\mathrm{cal}}$—obtained as the square root of the dark gray area in the spectrum—is used to determine the transfer function via Eq.\,(\ref{trans}).}
    \label{fig:cal}
\end{figure}

The ferromagnet with magnetic moment $\vec{\mu}$ is constrained by the applied bias field $\vec{B}_{\mathrm{bias}}$, leading it to undergo small oscillations around the field direction. Using the digital lock-in technique described in our previous work \cite{Ren} to probe the FMTO’s dynamic response, we previously demonstrated that the system exhibits two oscillation modes below 10 Hz: a pendulum mode and a torsional mode, as shown in Fig.\,\ref{fig:cal}. The torsional mode corresponds to rotations of the sensing magnet about its own symmetry axis. Because the torque generated by $\vec{B}_{\mathrm{bias}}$ is aligned with this axis, the squared eigenfrequency of this mode can be tuned by the bias current, exhibiting an approximately linear dependence, as shown in the inset of the left panel of Fig.\,\ref{fig:cal}. We select this torsional mode for axion detection, and its dynamics are governed by
\begin{equation}
    I\ddot{\theta}+I\gamma \dot{\theta}+I\omega _{0}^{2}\theta =\tau_s,
    \label{fmtoeom}
\end{equation}
where the resonant angular frequency $\omega_0$ is defined by $\omega_0\equiv\sqrt{\mu B_{\mathrm{bias}}/I}$, and $\gamma$ is the damping factor. $\tau_s$ is the projection of the signal torque onto the detector’s sensitive direction $\hat{\xi}$. The solution of Eq.\,(\ref{fmtoeom}) expressed in frequency domain is
\begin{align}
    \theta(f)=\chi(f) \tau_s(f)/I,
    \label{theta}
\end{align}
where $\tau_s(f)$ is the spectrum of the driving torque, $\chi(f) = 1 / \left( 4\pi^2 \left(f_0^2-f^2+i f f_0 / Q\right) \right)$ is the mechanical susceptibility of damped oscillator, where $Q=\omega_0/\gamma$ is the quality factor with $\omega_0=2\pi f_0$. The quality factor and resonance frequency in the present experiment are recalibrated as $Q = 39.2\pm 0.4$ and $f_0 = 5.00\pm 0.01 \,\mathrm{Hz}$, respectively.

The raw observable is the displacement of the optical spot on the imaging plane, recorded in pixel units as $Y(t)$, which traces the motion of the sensing magnet. Our analysis is performed in Fourier space on the magnetic PSD $S_B(f)$, obtained from the displacement PSD $S_Y(f)$ through a transfer function $\mathcal{T}(f)$. To determine $\mathcal{T}(f)$, we follow the calibration procedure established in Ref.\,\cite{Ren}. A sinusoidal magnetic field with known amplitude $B_{\rm cal}$ and frequency $f_{\rm cal}$ is applied in an off-resonant region, producing a calibration peak in the displacement spectrum (lower panel of Fig.\,\ref{fig:cal}). From the measured displacement amplitude $Y_{\rm cal}$ and the mechanical susceptibility $\chi(f)$, the transfer function is given by
\begin{align}
    \mathcal{T}(f)=\frac{B_{\text{cal}}^2}{2 Y_{\text{cal}}^2}\frac{\left\vert \chi(f_{\text{cal}})\right\vert^2}{\left\vert\chi(f)\right\vert^2}.
    \label{trans}
\end{align}
Here $Y_{\rm cal}$ is extracted as the square root of the integrated power under the calibration peak. The magnetic PSD is then obtained via $S_B(f)=\mathcal{T}(f)S_Y(f)$.

\section{Data Characterization and Statistical Analysis}

\begin{figure}[htbp]
    \centering
    \includegraphics[width=0.7\textwidth]{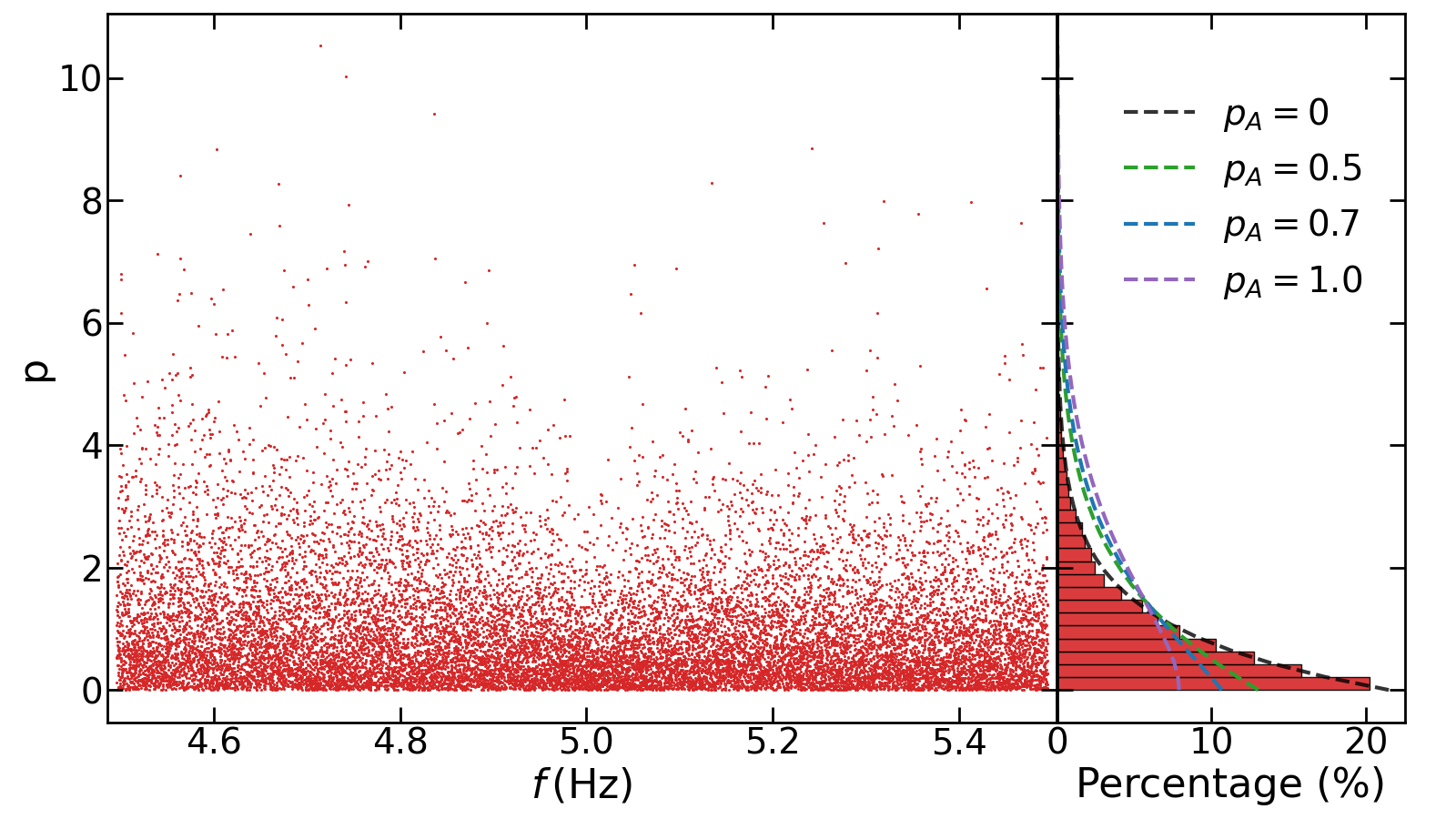}
    \caption{Measured power excess $p$ in each frequency bin together with its statistical distribution. The observed distribution is well described by a $\chi^2$ distribution with two degrees of freedom ($p_A = 0$). For comparison, the distributions Eq.\,(\ref{noncenchi2}) for $p_A=0.5,\,0.7,\,1.0$ are also shown.}
    \label{fig:psd_distribution}
\end{figure}

The axion search data were recorded continuously for 6 hours starting at 23:19. This observation window was deliberately chosen to avoid disturbances from daytime subway activity. Our prior characterization of the FMTO system indicates that subway activity can increase the ambient magnetic noise level by nearly an order of magnitude \cite{Ren}. Therefore, data acquisition was scheduled during nighttime hours to ensure optimal magnetic quiet conditions. For a set of measured data points only contributed by the noise, the power excess $p(f)\equiv S_B(f)/\bar{S}_B(f)$, defined as the PSD normalized by the expected noise baseline, follows a $\chi^2$ distribution with two degrees of freedom according to the central limit theorem \cite{1975ApJS...29..285G}. When a monochromatic signal such as the one we expect from the axion dark matter is present, the measured excess power in the bin containing the signal should be distributed according to a non-central $\chi^2$ distribution with two degrees of freedom and non-centrality parameter $p_A$ given by Eq.\,(\ref{ps}) \cite{1975ApJS...29..285G}, i.e.,
\begin{align}
    P_A(p; p_A) = e^{-(p + p_A)}I_0\left[2(p p_A)^{\frac12}\right],
    \label{noncenchi2}
\end{align}
where $I_0(z)=\sum_{k=0}^\infty ((1/4)z^2)^k /(k!)^2$ is the modified Bessel function of the first kind. When $p_A=0$, Eq.\,(\ref{noncenchi2}) reduces to the $\chi^2$ distribution with two degrees of freedom. In Fig.\,\ref{fig:psd_distribution}, we show the measured power excess p for each frequency bin along with its statistical distribution, which is consistent with the expectation for a noise-dominated spectrum.

To account for the stochastic variables of axion dark matter characterized by the amplitude $\alpha$, we need to consider the marginalized likelihood
\begin{align}
    \mathcal{L}(p;A)&=\int P_A(p; p_A(\alpha)) P_\alpha(\alpha) d\alpha \nonumber \\
    &=\frac{1}{1+2A}e^{-\frac{p}{1+2A}}.
    \label{likeli}
\end{align}
The random variable $\alpha$ follows the probability densities given in Eqs.\,(\ref{random}), and they take constant values over a coherence time of axion dark matter. Since each of them is statistically independent, their joint probability density is the product of their individual densities. In the absence of the axion signal, i.e., when $A = 0$, the likelihood Eq.\,(\ref{likeli}) reduces to an exponential distribution.

In addition to the statistical uncertainty arising from noise fluctuations, systematic uncertainties associated with calibration and fitting parameters also contribute to the total error budget. The major sources of such systematic uncertainties are summarized in Table.\,\ref{sigma}. For the lock-in–based calibration of the quality factor $Q$ and resonance frequency $f_0$, the uncertainty in $Q$ arises from the fitting error due to the finite number of frequency points in the sweep. In contrast, the fitting uncertainty of $f_0$ is smaller than the frequency step of $0.01\,\mathrm{Hz}$, and its uncertainty is therefore determined by the sweep step size. For the calibration signal, the uncertainty in the excitation frequency $f_{\mathrm{cal}}$ is set by the output accuracy of the signal generator. The uncertainty in the induced displacement amplitude $Y_{\mathrm{cal}}$ is determined from repeated measurements. The magnetic-field amplitude $B_{\mathrm{cal}}$ is obtained by calibrating the voltage-to-magnetic-field conversion factor of the drive coil via the current-amplifier setup, and its uncertainty dominates the systematic error budget. According to the error propagation formula, the resulting systematic uncertainty of the power excess in each frequency bin falls within the range $\sigma_{\mathrm{sys},p} \in [0.22, 0.26]$, which is smaller than the statistical uncertainty $\sigma_p \approx 1$ caused by noise fluctuations. To incorporate the effect of systematic uncertainties, a modified likelihood function can be considered as follows:
\begin{align}
    \mathcal{\tilde{L}}(p;A)&=\int \mathcal{L}(p+\delta p;A) P_G(\delta p) d\delta p \nonumber \\
    &=\frac{1}{1+2A}e^{\frac{-2(1+2A)p+\sigma_{\mathrm{sys},\,p}^2}{2(1+2A)^2}},
    \label{modlikeli}
\end{align}
where $P_G(\delta p)$ is the Gaussian distribution of $\delta p$ with zero mean and a standard deviation of $\sigma_{\mathrm{sys},\,p}$.

In order to perform our inferences, we define the likelihood ratio
\begin{align}
    \tilde{\lambda}(A) \equiv \frac{\mathcal{\tilde{L}}(p;A)}{\mathcal{\tilde{L}}(p;\hat{A})}.
    \label{likelira}
\end{align}
The parameter $\hat{A}$ is the maximum-likelihood estimator of $A$, maximising the likelihood given the measured value of the excess power $p(\omega)$. If $\hat{A}$ falls outside of its physical domain of $A\in [0,\infty)$, $\hat{A}$ is set to zero. We then construct the two-sided test statistic (TS) \cite{Cowan:2010js}
\begin{align}
    \tilde{t}_{A}\equiv -2\ln{\tilde{\lambda}}(A),
    \label{tA}
\end{align}
which is used to assess whether the null hypothesis, defined according to the likelihood in the numerator of Eq.\,(\ref{likelira}), is a good description of the observed data compared to the alternative hypothesis, defined as per the maximised unconditional likelihood. A value of 0 for $\tilde{t}_{A}$ indicates perfect agreement with the null hypothesis, with larger values indicating greater disagreement.

To reject the null hypothesis, we define a value for the TS that is deemed too extreme to have come from that model. Specifically, we consider the conditional probability distribution $f\left(\tilde{t}_A \vert A\right)$, which specifies the distribution of possible values of $\tilde{t}_A$ under the null hypothesis characterized by the parameter $A$. The p-value associated with the observed TS $\tilde{t}_A^{\text{obs}}$ is then defined as
\begin{align}
    p_{\rm val}^A \equiv \int_{\tilde{t}_A^{\text {obs}}}^{\infty} f\left(\tilde{t}_\kappa \vert A\right) \mathrm{d} \tilde{t}_A,
\end{align}
which tells us how probable it is for us to have observed a value of $\tilde{t}_A$ at least as extreme as $\tilde{t}_A^{\text{obs}}$. To compute the distribution $f\left(\tilde{t}_\kappa \vert A\right)$, we apply a Monte Carlo simulation strategy. For each value of $A$ to be tested, we generate $10^6$ pseudo-datasets according to the likelihood given in Eq.\,(\ref{likelira}). For each dataset, $\tilde{t}_A$ is calculated using Eq.\,(\ref{tA}), thereby obtaining the probability distribution $f\left(\tilde{t}_\kappa \vert A\right)$ of its values.

To evaluate the discovery significance, we examine the local p-value $p_{\rm val}^0$. For every measured power excess $p(\omega)$ in each frequency bin, we compute $\tilde{t}_A^{\mathrm{obs}}$ and its associated $p_{\rm val}^0$. The smallest local p-value obtained is $p_{\rm val}^0 = 3.3\times 10^{-5}$, which approaches $4\sigma$ threshold of $p_{\rm val}^0 \approx 6.3\times 10^{-5}$. However, because the search spans $N_p = 18{,}000$ statistically independent frequency bins, we must account for the look-elsewhere effect. The resulting global p-value is
\begin{align}
    p_{\rm global,val}^0 = 1 - (1 - p_{\rm val}^0)^{N_p} \approx N_p p_{\rm val}^0 = 0.59,
\end{align}
corresponding to a global significance of only $0.53\sigma$. Therefore, the observed excess is fully consistent with statistical fluctuations, and we do not claim a discovery; instead, we proceed to derive exclusion limits.

We proceed to compute our 90\% confidence level limits. We define the false positive rate as $0.1$ and find that the value of $A_{\mathrm{lim}}$, for which $p_{\rm val}^{A} = 0.1$ within each frequency bin. The resulting limit is shown by orange line in Fig.\,\ref{fig:constraints}. We then generate $10^6$ background-only pseudo-datasets according to Eq.\,(\ref{likeli}) with $A=0$. For each of these datasets, we find $A_{\mathrm{lim}}$, producing a distribution of $A_{\mathrm{lim}}$ values. Based on this, we derive the median value with $1\sigma$ and $2\sigma$ bands of $g_{aee}$ shown in Fig.\,\ref{fig:constraints}. We find excellent agreement between the results from our simulation and our derived data-driven limit, which closely follows the median limit and lies well within the $2\sigma$ band.

\section{Future Experiments}

\begin{figure}[htbp]
    \centering
    \includegraphics[width=0.45\textwidth]{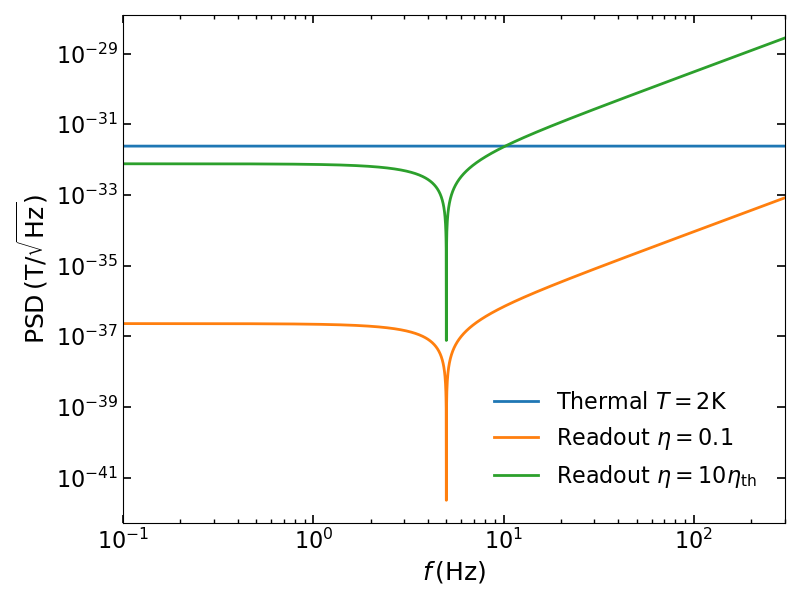}
    \caption{Thermal and readout noise PSD calculated for $T = 2\,\mathrm{K}$ and $Q = 10^5$.}
    \label{fig:noise}
\end{figure}

By improving the magnetic shielding design to suppress environmental magnetic noise and upgrading the suspension system to reduce mechanical vibration noise, the dominant noise contributions in the system can be expressed as
\begin{align}
    S_{B}^{\mathrm{tot}}=S_{B}^{\mathrm{th}}+S_{B}^{\mathrm{imp}}+S_{B}^{\mathrm{ba}},
    \label{totn}
\end{align}
where $S_{B}^{\mathrm{th}}$ is the environmental thermal magnetic noise PSD, which is related to the torque PSD $S_{\tau}^{\mathrm{th}}$ through
\begin{align}
    S_{\tau}^{\mathrm{th}} = 4 k_B T I \frac{\omega_0}{Q},\  S_B^{\mathrm{th}}= S_\tau^{\mathrm{th}}/\mu^2,
    \label{thermal}
\end{align}
where $\omega_0$ is the resonant angular frequency, $Q$ is the quality factor, $T$ is the environmental temperature, $\mu$ is the magnetic moment of the sensing magnet, and $I$ is the moment of inertia of the oscillator. $S_{B}^{\mathrm{imp}}$ and $S_{B}^{\mathrm{ba}}$ together constitute the readout noise, which originate from the imprecision noise $S_{\theta}^{\mathrm{imp}}$ in the measurement of the deflection angle and the back action noise $S_{\tau}^{\mathrm{ba}}$ exerted by the laser on the oscillator during the measurement, respectively. These two types of noise satisfy the uncertainty relation
\begin{align}
    S_{\theta}^{\mathrm{imp}}S_{\tau}^{\mathrm{ba}} = (1/\eta) \hbar^2,
\end{align}
where we introduce $\eta\in [0,1]$ to characterize the measurement efficiency, and $\eta=1$ corresponds to the standard quantum limit. For the readout noise, we have $S_{B}^{\mathrm{imp}}+S_{B}^{\mathrm{ba}}\geq 2\sqrt{S_{B}^{\mathrm{imp}} S_{B}^{\mathrm{ba}}}$, and the minimum value can be achieved by tuning the readout apparatus such that $S_{B}^{\mathrm{imp}} = S_{B}^{\mathrm{ba}}$. Therefore, based on $S_{B}^{\mathrm{ba}}=S_{\tau}^{\mathrm{ba}}/\mu^2$ and $S_{B}^{\mathrm{imp}}=I^2 \vert\chi(\omega)\vert^{-2} S_{\theta}^{\mathrm{imp}} /\mu^2$, we obtain
\begin{align}
    S_{B}^{\mathrm{imp}}+S_{B}^{\mathrm{ba}}=2\eta^{-\frac12}\hbar \mu^{-2} I \vert\chi(\omega)\vert^{-1}.
    \label{readout}
\end{align}
Under fixed temperature $T$ and measurement efficiency $\eta$, the effective sensitivity bandwidth limited by thermal noise is determined by solving for the frequency range where $S_{B}^{\mathrm{th}} > S_{B}^{\mathrm{imp}}+S_{B}^{\mathrm{ba}}$. Therefore, in a search experiment that scans axion masses by tuning the resonance frequency, the scanning efficiency can be improved by adjusting the quality factor $Q$ to maximize the thermal-noise–dominated sensitivity bandwidth \cite{Chaudhuri:2018rqn}. For the FMTO system, the feasibility of tuning the quality factor $Q$ during the mass-scanning process remains an open question; thus, we treat it as a constant in our estimation. For the measurement efficiency $\eta$, we require that at frequencies far below the resonance, the total system noise is dominated by thermal noise. This condition sets a lower bound on $\eta$, denoted as $\eta_{\mathrm{th}}$
\begin{align}
    \eta_{\mathrm{th}}=\left(\frac{\hbar \omega_0 Q}{2 k_B T}\right)^2.
\end{align}
We expect to achieve $\eta > \eta_{\rm th}$ in future measurements, so that the sensitivity at and below resonance becomes thermal-noise limited. In Fig.\,\ref{fig:noise}, we show the readout noise PSDs for $\eta = 10\eta_{\mathrm{th}}$ and $\eta = 0.1$, together with the thermal-noise PSD at $T = 2\,\mathrm{K}$ and $Q=10^5$.

When $T_{\mathrm{mea}}>T_{\mathrm{coh}}$, the monochromatic approximation for the axion field given in Eq.\,(\ref{unaf}) no longer holds. The total measurement time $T_{\mathrm{mea}}$ can instead be divided into multiple segments, each of duration equal to the axion coherence time $T_{\mathrm{coh}}$. These segments correspond to statistically independent measurements, reducing the overall noise level by a factor of $\sqrt{N}$, where $N = T_{\mathrm{mea}} / T_{\mathrm{coh}}$. This is equivalent to replacing $T_{\mathrm{mea}}$ in the signal PSD Eq.\,(\ref{signalPSD}) with an effective observation time $T_{\mathrm{eff}}$, given by
\begin{align}
    T_{\mathrm{eff}}=\left\{\begin{array}{l}
    T_{\mathrm{mea}}\ \ \ \mathrm{if}\ T_{\mathrm{mea}}<T_{\mathrm{coh}},  \\
    \sqrt{T_{\mathrm{mea}} \cdot T_{\mathrm{coh}}}\ \ \ \mathrm{if}\ T_{\mathrm{mea}}>T_{\mathrm{coh}}.
    \end{array}\right.
    \label{teff}
\end{align}

To estimate the projected sensitivity to the axion–electron coupling strength, we consider the condition where the signal-to-noise ratio satisfies $\mathrm{SNR} = 1$. According to Eqs.\,(\ref{signalPSD}) and (\ref{totn}), the expected limit on $g_{aee}$ can be obtained by solving the relation $S_A(g_{aee}) = S_B^{\mathrm{tot}}$, where the total measurement time $T_{\mathrm{mea}}$ is replaced by the effective integration time $T_{\mathrm{eff}}$ as defined in Eq.\,(\ref{teff}). The random axion-field amplitude is taken to be its mean value $\alpha=2$, and the angle between the axion wind and the detector’s sensitive axis is fixed to the sidereal-day average $\cos^{2}\psi = 0.22$, which is larger than the 18{,}000\,s value $\cos^{2}\psi = 0.061$ used in the analysis of the present data set.

\section{Shielding of Axion Wind}
\label{sec:shielding}

In experiments employing magnetic shielding, interactions between external magnetic fields and electrons in the shielding material can induce an internal response field $\vec{B}_{\mathrm{in}}$, which may affect axion detection. Ref.\,\cite{JacksonKimball:2016wzv} argued that $\vec{B}_{\mathrm{in}}$ is oriented opposite to the axion-induced effective field $\vec{B}_a$, potentially suppressing the observable signal and thereby weakening the constraints derived in axion–electron coupling experiments using magnetic shielding. Notably, several existing searches for ultralight axion dark matter via axion–electron interactions—including the Torsion Pendulum and earlier comagnetometer experiments \cite{Terrano:2019clh,Bloch:2019lcy}—have not incorporated the effect of magnetic shielding in their data analyses. Likewise, a number of exotic-interaction experiments, in which sources and sensors are magnetically shielded from each other, have reported limits on exotic field–electron couplings without accounting for shielding-induced attenuation \cite{Xu:2025lly,Lee:2018vaq,Almasi:2018cob,Wu:2021flk}. For this reason, we do not include shielding effects in the main text.

However, to assess the maximal possible impact, we provide here a conservative estimate based on the arguments of Ref.\,\cite{JacksonKimball:2016wzv}, which suggests that the response of the permalloy magnetic shielding material to the axion field is similar to its response to a conventional DC magnetic field, with the key difference being that only the innermost layer of the shielding will affect the signal of $\vec{B}_a$. In our setup, since the sensor magnet is much smaller than the innermost shielding cylinder whose thickness is much smaller than its radius, the shielding can be approximated as an infinitely long cylindrical shell. The attenuation factor is then
\begin{align}
    G \approx \frac{2a}{\mu_r w} \approx 1.7\times 10^{-3},
    \label{shielding}
\end{align}
where $a=82.5\,\mathrm{mm}$ is the radius, $w=1\,\mathrm{mm}$ is the thickness, and we have taken the maximal relative permeability of permalloy, $\mu_r \approx 10^5$ \cite{coey2010magnetism}. This suggests that the experimental constraints could be weakened by up to three orders of magnitude—comparable to the shielding factor $\eta_s \approx 1/300$ reported in Ref.\,\cite{JacksonKimball:2016wzv} for ordinary magnetic fields in the range $10^{-6}\,\mathrm{G}$–$10^{-3}\,\mathrm{G}$. It is therefore reasonable to expect that other axion–electron coupling experiments employing magnetic shielding may experience a suppression of similar magnitude. Notably, Ref.\,\cite{JacksonKimball:2016wzv} suggests that superconducting magnetic shielding operates via electron orbital angular momentum rather than spin, and therefore does not induce $\vec{B}_{\mathrm{in}}$ that could suppress axion signals. Consequently, the projected low-temperature experimental limits shown in Fig.\,\ref{fig:prospect} are not expected to be affected by shielding effects.

A complete theoretical and experimental understanding of how axion-induced effective fields interact with magnetic shielding materials, however, is still lacking. One fundamental distinction is that the axion-induced pseudo-magnetic field is divergence-full, unlike real magnetic fields governed by Maxwell’s equations, potentially invalidating the scalar magnetic potential formalism commonly used in shielding analyses. Moreover, because $\vec{B}_a$ is extremely weak, it is unclear whether the magnetic domains inside an overall unmagnetized shielding material can respond efficiently. For ordinary magnetic fields, it is well established that the permeability of soft ferromagnets drops sharply in the weak-field limit \cite{Gohil:2020ywu,sun2022method}, raising further questions regarding the shielding response to axion-induced fields.

\end{document}